# Isomeric and hyperdeformed states at high spin for light nuclei

A. Boucenna, S. Madjber and A. Bouketir

*Département de Physique, Faculté des Sciences*

*Universite Ferhat Abbas, 19000 Setif, Algeria*

We present a macroscopic model for the energy of rotation nuclei which has several refinements relative to the rotating liquid drop model. The most important features are the inclusion of the shell correction and using a new family of triaxial compact and creviced shapes to describe transition of the system from one sphere to a two separated spheres. With this model we calculate the properties of point of equilibrium corresponding to nuclear ground states, isomeric states and fission saddle-points. The model is applied to the nuclei $^{56}$Ni, $^{48}$Cr and $^{80}$Zr. The results obtained allowed to explain the light compound nuclear system fission phenomena and predict the existence of hyperdeformed states at very high spins.

PACS numbers : 25.70.Jj, 25.70Gh, 24.60.Dr

## I- INTRODUCTION

In recent years, a number of experimental and theoretical studies have been made to understand the decay of light compound nuclear systems formed through heavy-ion reactions [1, 2, 3, 4, 5, 6, 7, 8, 9, 10, 11, 12 ]. The strong resonance-like structures observed in elastic and inelastic excitation functions of $^{24}$Mg + $^{24}$Mg [13] and $^{29}$Si + $^{29}$Si [5] have indicated the presence of shell stabilized, highly deformed configuration in $^{48}$Cr and $^{56}$Ni Compound systems, respectively. The systematic of fusion followed by fission in the light compound nuclei region is well established [3, 7 ]. The measurements of fission like yields where the $^{48}$Cr system is populated using the asymmetric-mass $^{36}$Ar + $^{12}$C and $^{20}$Ne + $^{29}$Si reactions [10] illustrate the important role of statistical fission in the general behavior of this system. Complete fusion of two heavy " light " ions at energies near the coulomb barrier leads to a light compound nucleus having a high excitation energy and a high angular momentum. Several authors have suggested the importance of fission decay in these light systems [1, 3, 14, 15, 16, 17, 18, 19, 20, 21]. In studying the fission like yield of the $^{16}$O + $^{16}$O reaction, it is found that many of the character features of this yield for the two systems can be understood in terms of fusion-fission model [22, 23]. One assumes the formation of a fully equilibrated compound nucleus that subsequently breaks apart into two fragments.

The origin and importance of the nuclear deformations have captured the attention of nuclear surface and intermediate energy heavy ion physicist for half a century now. An exciting resurgence is brought on the development of very powerful experimental facilities such as Eurogam and Gammasphere which allowed the investigations of nuclei in the intermediate energy domain (20-100 MeV/u). These nuclear systems are formed in heavy ion collisions (light nuclei) with high angular momentum. The discovery of superdeformed actinide isomers [24] was the first observation of deformed nuclear shapes. Later on, superdeformed ground states in $^{72}$Ge, $^{100}$Sr and $^{72}$Kr were found [ 25, 26]. The study of superdeformed states at high spins with the detection of cascade in $^{192}$Ce followed after by a rotational band [27] in $^{152}$Dy. Hyperdeformed nuclear shapes have also been seen. Indeed, Hyperdeformed fission isomers have been detected in a third well in Th nucleus [28] and evidence for the hyperdeformed prolate nuclear shape at high angular momentum has also been observed [29] in $^{152}$Dy.

Theoretical studies of compound systems are based on the rotating liquid drop model [30] to define the fission path and the equilibrium state configuration needed to know the existence of fused nuclei. Effects resulting from the finite range of the nuclear interaction and the diffuseness of the light systems clear surface have an important influence on the macroscopic energies calculated the light systems. Theoretically the entrance of the compound system in the deformation path [31] is first attempt to explain the experimental data. When the nucleus has a high angular momentum, fission barrier height decreases and gives the highest probability of fission decay comparatively with the other decay channels.

In the theoretical studies of Mustafa et al. [32] and Sierk [23] the finite range effect and the finite surface diffuseness effects are introduced to compute potential energy and to get real values in light nuclei region. The purpose of the present work is to find the isomeric and hyperdeformed states for light compound nuclei. The macroscopic approach is based on the Macroscopic Model of Rotating Nuclei [33] where the triaxial, compact and creviced shapes parameterization is used to describe the transition of the system from one sphere to two separated spheres and vice-versa in order to study the fission and fusion paths and explain the phenomena associated with ion interactions. The deformation energy of a nucleus is determined within the liquid drop model including the finite range effects in the nuclear surface energy and finite surface diffuseness effects in the coulomb energy and rotational moment of inertia. Since the shell correction is also taken to influence the fission decay probabilities, the shell correction is included.

**II- INGREDIENTS OF THE MODEL**

The rotating liquid drop model [30] considers the nucleus as an incompressible fluid, with a constant charge density and a sharp surface, which rotates as a rigid body. Nuclear surface energy, repulsive coulomb energy and rotating energy are the main components of the energy in this model. In the liquid drop model, surface nuclear energy suffers from several deficiencies to describe real nuclei. The proximity effects have an important part in nuclear surface energy of compact and creviced shape, near or at the saddle point of light nuclei or in the heavy ion fusion process. Krape, Nix and Sierk [22, 34] replaced the surface energy of the liquid drop model by the Yukawa nucleon energy to modify the effects of the finite range of the nuclear forces and the finite surface of a real nucleus. The range of Yukawa functions does not agree with experimental data of heavy ion interactions. Krape et al. [35] resolved these discrepancies. In the following section we describe the model and the procedure used in our calculations of the properties of the rotating nuclei. This model is based on the Macroscopic Model of Rotating Nuclei of Sierk [33] which incorporate a new shape parametrization and a shell correction.

**A. Shape parameterization**

To study the fission process in the phenomenological models, the spatial deformation of the surface represents the first step. The saddle points of the light nuclei are predicted by the liquid drop model as a configuration close of two touching spheroids separated by a relatively narrow neck region [3] initially formed by fusion from "heavy" ions. As the system passes over the saddle point, towards greater deformation of the compound system, the neck eventually breaks at the scission point. The shape parameterization, discussed by Nix [36], consists of three connected quadratic surface of revolution. The previous works [37, 31] predict that the fission of light nucleus lead to compact and creviced shapes. In attempt to explain the fission of light nucleus, the existence of triaxial ground states and the existence of hyperdeformed nuclei we have generalized the elliptic lemniscatoids shape parameterization which permits the description of transition system from two separated spheres to one body, and defined in cylindrical coordinates ($\rho$, $\phi$, z) by:

$$a^2\rho^2 + c^2z^2 = (\rho^2 + z^2)^2 \tag{35}$$

and in spherical coordinates (R, θ, φ) the shape is given by:

$$R^2(\theta) = a^2\sin^2\theta + c^2\cos^2\theta$$

where a is the neck radius and c is the nucleus half elongation. This family forms the elliptic lemniscatoids, which are obtained by inversion of the axially symmetric ellipsoids and generalized to the left-right asymmetry. If we assume a volume conservation during the deformation, the two parameters s = c/a and $\beta$ = $R_2/R_1$, where $R_1$ and $R_2$ are the radii of the fragments, are sufficient to define completely the shape. For symmetric left right shape ($\beta$ = 1) decreasing s from one to zero gives continuous variation of the shape from a sphere to two touching spherical fragments with a deep neck formation. The equation (35) can be written in cylindrical coordinates ($\rho$, φ, z) as:

$$\rho^2 = \frac{1}{2}\left(s^2c^2 - 2z^2 + c\sqrt{s^4c^2 + 4z^2(1-s^2)}\right) \tag{36}$$

and in cartesien coordinates (x, y, z) we can write :

$$x^2 + y^2 = \frac{1}{2}\left(s^2c^2 - 2z^2 + c\sqrt{s^4c^2 + 4z^2(1-s^2)}\right) \tag{37}$$

This shape is axially symmetric. In order to describe triaxial shapes, we are replacing in Eq. (37) $x^2 + y^2$ by $x^2 + \alpha y^2$, where α is a new parameter which break the axial symmetry.

The surface of triaxial nucleus is defined in cylindrical coordinates by :

$$\rho^2 = \frac{1}{2(1+(\alpha-1)\sin^2\phi)}\left(s^2c^2 - 2z^2 + c\sqrt{s^4c^2 + 4z^2(1-s^2)}\right) \tag{38}$$

The φ coordinate appears explicitly in the expression of ρ, and which justifies the triaxiality of our parametrization. In the symmetry left-right case two dimensionless parameters s = a/c and α (α≠0) are sufficient to define completely the shape sequence, when in the case of left-right asymmetry, it is necessary to add a third parameter to define completely the shape and the shape parametrization is given by :

$$\rho^2 = \frac{1}{2(1+(\alpha-1)\sin^2\phi)}\left(s_1^2c_1^2 - 2z^2 + c_1\sqrt{s_1^4c_1^2 + 4z^2(1-s_1^2)}\right) \quad z>0 \tag{40}$$

$$\rho^2 = \frac{1}{2(1+(\alpha-1)\sin^2\phi)}\left(s_2^2c_2^2 - 2z^2 + c_2\sqrt{s_2^4c_2^2 + 4z^2(1-s_2^2)}\right) \quad z<0 \tag{40}$$

The three dimensionless parameters $s_1 = a/c_1$, $s_2 = a/c_2$ and $\alpha$ are sufficient to define shape completely. The Ratio of the radii of the two colliding nuclei:

$$\beta = \frac{R_1}{R_2} \qquad (41)$$

allows us to connect $c_2$ with $c_1$ and to introduce explicitly the asymmetry degree of freedom all along the fusion path. The simplest way to connecting $c_2$ with $c_1$ is obtained through the following quadratic expression:

$$c_2^2 = s_1^2 c_1^2 + (1 - s_1^2)\beta^2 c_1^2 \qquad (42)$$

such that :

$$s_2^2 = \frac{s_1^2}{s_1^2 + (1 - s_1^2)\beta^2} \qquad (43)$$

The shape sequences are as followed: For $\alpha = 1$ and $s_1$ increases from 0 to 1, the shape varies continuously from two touching spherical nuclei to a one spherical nucleus. For $0<s<1$ and $0<\alpha<9$, the shape sequence describes the triaxial, compact and creviced shapes. The use of this shape parameterization allows us to study, point by point, the path fission without calculating high order derivative deformation energy terms and to explain the occurrence of hyperdeformed nuclei at very high angular momentum.

**B - The shell effects**

The heavy ion resonance phenomenon observed in some light nuclear systems may reflect a strong shell correction to the fission potential energy surface [14, 15]. In our calculations the shell effects are introduced using a semi-empirical Myers's method [38 ]:

$$E_{shell} = E_{shell}^{(0)}(1 - 2\varepsilon^2)e^{-\varepsilon^2} \qquad (29)$$

where $E_{shell}^{(0)}$ is the spherical shell corrections which is given by :

$$E_{shell}^{(0)} = f_1 \left[ \frac{F(N) + F(Z)}{(\frac{1}{2}A)^{2/3}} - f_2 A^{2/3} \right] \qquad (29)$$

Where N, Z and A are the neutron, proton and nucleon numbers respectively, and :

$$F(N) = q_i(N - M_{i-1}) - \frac{3}{5}(N^{5/3} - M_{i-1}^{5/3}) \quad \text{for} \quad M_{i-1} < N < M_i \tag{30}$$

The quantities $q_i$ are defined as :

$$q_i = \frac{3}{5} \frac{M_i^{5/3} - M_{i-1}^{5/3}}{M_i - M_{i-1}} \tag{31}$$

Where :

$$M_i = 2, 8, 14, 29, 50, 82, 126 \tag{32}$$

$f_1$ and $f_2$ are two adjustable coefficients :

$$f_1 = 5.8 \text{ MeV} \quad \text{and} \quad f_2 = 0325 \tag{33}$$

The parameter $\varepsilon$ is the deviation of the surface from a sphere :

$$\varepsilon^2 = -\frac{\delta R^2}{a^2} \tag{34}$$

The range has been chosen to be 0.32 $r_0$, where $r_0$ = 1.2499 [37]. Using this approach, shell effects only play a role near the ground state of the compound nucleus and not at the saddle-point, since shell effects are properties of valence nucleons and the orbitals of which are perturbed by the strong nuclear potential.

**C - Deformation energy**

The computation of the deformation energy represents the most important operation to determine the equilibrium configuration and the study of the fission. The deformation energy $E_d(N, Z, P_{def})$ is a difference between the deformed nucleus energy E and the spherical nucleus energy $E^{(0)}$ :

$$E_d(N, Z, P_{def}) = E - E^{(0)} \tag{12}$$

Where N, Z are the neutron, proton numbers and $P_{def}$ are the parameters (s1, β, γ). γ is the deformation parameter defined by:

$$\gamma = \frac{r}{R_0}$$

Where $R_0$ is the spherical nuclear radius and r is the distance between the centers of the fragments. E and $E^{(0)}$ are given by :

$$E = E_V + E_s + E_c + E_R + E_{shell}$$

And :

$$E^{(0)} = E_V + E_s^{(0)} + E_c^{(0)} + E_{shell}^{(0)}$$

Where $E_s^{(0)}$, $E_c^{(0)}$, $E_{shell}^{(0)}$ are the surface, coulomb and shell energies of the spherical nucleus, $E_V$ is the volume energy and $E_s$, $E_c$, $E_R$, and $E_{shell}$ are the surface, coulomb, rotational and shell energies of rotating nucleus.

**III – RESULTS AND DISCUSSION**

The fission process in light nuclear systems can be successfully described using the same transition-state formalism that has become the standard description of fission in heavier systems. The development of saddle-point calculations in light system was significantly delayed from comparable development for heavy system fission because of the difficulty in accounting for the finite range and diffuse nuclear surface effects that strongly influence the macroscopic energies of these systems. In light nuclei, the saddle-point shapes correspond to two deformed spheroids separated by a well-developed neck region, with the surfaces of the two spheroids coming within close proximity of one another. The development of the finite range-range model [23], has made it possible to extend the saddle-point calculation to very light systems.

In this section we present the results of our calculations of path fission, the maximum angular momentum which supported by the nuclei along Green's approximation to the line of β-stability and the prediction of isomeric states and hyperdeformed nuclei.

We have evaluated both the surface, coulomb and rotational energy by transforming the integrals into surface integrals. The integrands were then transformed to stretched cylindrical coordinates (ρ, ϕ, z) and integrated over z and ϕ by means of a sixteen point Gauss-Legendre quadrature formula in order to reach high enough numerical accuracy within reasonable computing time.

The fission path curve gives the deformation energy $E_d(N, Z, P_{def})$ versus the deformation parameter $\gamma = \dfrac{r}{R_0}$. Our shape parameterization allows us to study, point by point, the path fission without calculating high order derivative deformation energy terms. For all non-rotating β-stable nuclei with A less than 300, the ground state is a sphere. like it was reported by [23], we distinguish three regions for the saddle point shape For A<220 the saddle point configuration is a two separated sphere system where the distance between these spheres increases when A decreases. For 220 ≤ A ≤ 245 these spheres approach until they touch each other and present the saddle point shape. For A ≥ 246 the saddle point shape is a body deep crevice. In the rotating case the barrier fission value diminish by centrifuge force effect resulting from the rotation of the nucleus. The ground state which is initially spherical become deformed into nearly spherical oblate shape with weak crevice at small values of angular momentum. When values increase the ground state tends to a prolate shape with deep crevice.

Figure 1 shows the calculated fission barrier values of β-stable nuclei. These barriers are less than those calculated from liquid drop model. This decreasing of fission barriers is a consequence of the finite range effects of nuclear force and surface diffuseness. These effects have an important part in decreasing fission barriers particularly when we use a creviced surface parameterization as shown in figure 2.

In table 1, we report the experimental values of the fission barriers height for the β stable nuclei. These values are compared to those calculated by the axial shape sequence. We also compare our results concerning some nuclei with experimental data and the prediction of other models. As example for the $^{109}$Cd nucleus, for which the shell effects are negligible, our calculations give a fission barrier height value 35.44 MeV, which is very close to the experimental fission barrier height 34.0 MeV. This interesting result predicts the triaxially ground state shapes for the $^{109}$Cd nucleus.

In figure 3 we show the variation of the fission barrier versus the angular momentum for A = 50, 100 and 200 β-stable nuclei. When increasing the angular momentum, fission barrier diminishes and becomes equal zero for the critical angular momentum $L_{cr}$. The nucleus cannot support upper angular momentum. In figure 4, we show the calculated values of the critical angular momentum for β-stable nuclei versus the mass number A. Our calculations give maximal angular momentum, $L_{max} = 135\,\hbar$ which occurs at A = 190. This value agrees with $L_{max} = 130\,\hbar$ at A=190 Royer value [39]. The maximum value of critical angular momentum supported by the nucleus, in the rotating liquid drop model, is $100\,\hbar$ and in ref. [39] and Mustafa et al. give $L_{max} = 75\,\hbar$ at A=140 [32]. This obtained very high maximal angular momentum shows the important role of the used shape surface in the prediction of the maximum angular momentum supported by the nucleus and shows the possibility of the formation of hyperdeformed states at very high spins.

One of the most important developments in nuclear structure physics was the prediction and observation of superdeformed shapes at high angular momentum. Calculations based on the cranking Strutinsky method with a deformed Wood-Saxon potential (41) had predicted the existence of hyperdeformed nuclear states but not so high spins. In particular, rotational bands built upon hyperdeformed shape with axis ratios around 3:1 are predicted to become yrast at spins as low as $70\,\hbar$, and could therefore be populated in heavy-ion fusion reactions. Hyperdeformation has been observed in $^{152}$Dy [29] where ridges were found in two-dimensional γ-γ coincidence spectra witch correspond to a high moment of inertia of $130\,\hbar^2$ MeV$^{-1}$, suggesting the existence of a strongly deformed prolate shape. This observed ridge structure occurs for a very high spin range about 75 - 95 $\hbar$. Our parametrization allow us to explain the occurrence of hyperdeformed nuclei at very high angular momentum. Calculations show that the nuclei can populate the very high angular momentum states and predict the angular momentum between $70\,\hbar$-$110\,\hbar$, for $^{152}$Dy. These values agree with experimental values [29], and with previous calculation values [31]. Figure 5 shows the region of the hyperdeformation for β–stable nuclei predicted by our calculations. The upper and lower limits of the angular momentum $L_I$ and $L_{II}$ are obtained assuming that the fission barrier is up than 0.5 MeV.

## $^{48}$Cr nucleus

While studying the strong resonance behaviors found in excitation function of the $^{24}$Mg+$^{24}$Mg [13] elastic and inelastic channels, a significant non-resonant background yield was discovered in the energy spectra of these channels at higher excitation energies. Measurements were shown to be consistent with the fission decay of the compound system. To explore the relationship of the different reaction mechanisms influencing the binary decay yields of the $^{48}$Cr compound systems, the yields of the reactions, leading to the $^{48}$Cr compound nucleus, have been studied [7, 7, 41, 42, 14, 15]. In calculations of the shape-dependant potentiel-energy surfaces at high angular momenta (16-$40\,\hbar$) for the $^{48}$Cr nucleus [43], a strong superdeformed configuration is predicted that correspond to aligned arrangement of two touching and highly deformed $^{24}$Mg nuclei. The fission path for $^{48}$Cr nucleus, calculated with a non creviced shape parameterization does not shows any structure which one can associate it to an isomeric state. Figure 6 shows the fission path for the $^{48}$Cr calculated with an axial creviced shape (α = 1) for L = $35\,\hbar$. For L<$36\,\hbar$, there is no structure and the fission path remains unchanging. A second peak appears at L = $36\,\hbar$, before the scission point at $\frac{r}{R_0}=1.538$, correspondig to an isomeric state, (figure 7). This result is in good agreement with the Wusmaa and Dudek works [40, 44] concerning the $^{24}$Mg + $^{24}$Mg reaction which produced a resonance at L = $36\,\hbar$. This superdeformed configuration is a candidate to become yrast at around spin $34\,\hbar$, in the high excitation energy region which corresponds to where the quasi-molecular resonances have been observed. At L = $37\,\hbar$ this second peak moves to $\frac{r}{R_0} = 1.64$ after the scission point, and the corresponding state is a superdeformed state. Figure 8 shows the coutour energy $E_d(s_1, \alpha)$ – the fission path – for the $^{48}$Cr for L = 35 $\hbar$ calculated with a triaxial crevaced shape ($\alpha \neq 1$). For L<36 $\hbar$, there is no structure and the fission path remains unchanging. A second minimum appears at L = 36 $\hbar$, after the scission point, at $\alpha$ = 1.6, correspondig to an isomeric state, (figure 9). A second minimum, also appears at L = 37 $\hbar$, after the scission point. But the most important result is that the $^{48}$Cr nucleus stays stable for a very high angular momentum (L = 60 $\hbar$) corresponding to 120 MeV excitation energy (figure 10) . And the hyperdeformed states for the $^{48}$Cr compound nucleus appear for $25\hbar \langle L \langle 40\hbar$ for both cases symmetric and asymmetric.

## $^{56}$Ni nucleus

The $^{56}$Ni compound nucleus has been explored through multiple entrance channels [5, 16, 18]. The yields of the $^{16}$O + $^{40}$Ca, $^{29}$Si + $^{29}$Si, $^{32}$S + $^{24}$Mg reactions are consistent with a fusion-fission reaction mechanism. The $^{56}$Ni is doubly magic nucleus, therefore very linking, where the shell corrections are very important for non-rotating nucleus. The fundamental state is spherical. Figure 11 shows the fission path for the $^{56}$Ni calculated with for L = 0 $\hbar$, it appears a second little peak near the spherical state at $\frac{r}{R_0} = 0.88$. As L increases, the fundamental state becomes quasi spherical. For L < 42 $\hbar$, the fission path for the $^{56}$Ni calculated with an axial creviced shape ($\alpha$=1) remains still unchanging (there is only one minimum on the contour energy). At L = 42 $\hbar$, a second minimum is observed at $\frac{r}{R_0} = 1.5874$, which is considered as an isomeric state, figure 12, and it explains the fusion fission phenomena and the orbiting mechanism observed in Sanders experimental works [33]. Figure 13 shows the energy contour $E_d(s_1, \alpha)$ for the $^{56}$Ni calculated with a triaxial crevaced shape ($\alpha \neq 1$) for L = 42 $\hbar$. Two minimum are observed, after the scission point at $\alpha$ = 1.7, corresponding to two isomeric states situated after the scission point. For the $^{56}$Ni nucleus, Our calculations predict hyperdeformed states for $20\hbar < L < 50\hbar$ (figure 14). These hyperdeformed states are not yet observed experimentally.

## $^{80}$Zr nucleus

Evans et al. [45] have been studied the $^{40}$Ca+$^{40}$Ca reaction at $E_{c.m.}$ = 197 MeV and 231 MeV. The $^{80}$Zr compound nucleus, populated through the $^{40}$Ca + $^{40}$Ca reaction, is found to have either a very flat distribution of barrier energies as a function of mass asymmetry or slight lower barriers for the symmetric mass configuration. The fundamental state of the $^{80}$Zr is spherical. When shell effects is include this state becomes quasi spherical. The fission path for the $^{80}$Zr calculated with an axial creviced shape ($\alpha$ = 1) for L < 59$\hbar$ remains unchanging (there is only one minimum on the contour energy). At L = 59$\hbar$, a structure is observed which is considered as an isomeric state, near the touching point (figure 15). Figure 16 shows the coutour energy $E_d(s_1, \alpha)$ for the $^{80}$Zr for L = 59$\hbar$ calculated with a triaxial crevaced shape ($\alpha \neq 1$). We do not observe minimum which it can be correspond to an isomeric state. Finally for the $^{80}$Zr nucleus, our calculations predict hyperdeformed states for 20$\hbar$ < L < 50$\hbar$ (figure 17). These hyperdeformed states are not yet observed experimentally.

## IV- CONCLUSION

Calculations based on the liquid drop model including a finite-range nuclear force and a diffuse nuclear surface with a triaxial, compact and creviced shape sequence predict the existence of isomeric and hyperdeformed states in compound light nuclei at very high spins. The calculated fission barriers are lower for light and medium mass nuclei than those found by other models and the critical angular moment $L_{BF}$, supported by the nuclei, is found higher for 120 ≤ A ≤ 250 and lower for A ≤ 80 than ones reported by other authors [32, 23]. These results are in good agreement with those of the fusion-fission process and with those of Royer [31]. This study explain the resonance phenomena observed in heavy ion interactions in term of isomeric states which are related to the second well in the fission path. Finally, the $^{48}$Cr, $^{56}$Ni and $^{80}$Zr have been particularly studied and isomeric and hyperdeformed states for these nuclei have been predicted.

**Table 1 :** Experimental and computed fission barrier values

$B_f(exp)$ : experimental values, $B_f(1)$ : this work, $B_f(2)$ : ref. 37, $B_f(3)$ : ref. MO76

**Table 1**

| Nucleus | $B_f(exp)$ | $B_f(1)$ | $B_f(2)$ | $B_f(3)$ |
|---|---|---|---|---|
| $^{50}$V | 29.5 1.17 | 30.6 | 31.3 | |
| $^{109}$Cd | 34.0 | 35.44 | 40.8 | 73.8 |
| $^{149}$Eu | 32.5 2 | 34.20 | 33.7 | 40.4 |
| $^{160}$Dy | 27.14 | 30.22 | 33.9 | 69.2 |
| $^{173}$Lu | 27.1 | 29.60 | 30.0 | 30.04 |
| $^{212}$Po | 19.7 | 21020 | 21.4 | 58.8 |
| $^{232}$Th | 6.0 | 12.56 | 8.3 | 45.7 |
| $^{234}$U | 6.0 | 11.21 | 6.9 | 46.3 |
| $^{240}$Pu | 5.7 | 10.31 | 5.7 | 45.6 |
| $^{246}$Cm | 4.7 | 8.82 | 4.73 | 41.7 |
| $^{250}$Bk | 5.58 | 6.60 | 4.2 | |

**Figure Captions**

**Figure 1.** Calculated fission barrier values of β-stable nuclei versus mass number A.

   ———  this Work

   • • • • •  Ref. 32

   – · – · – ·  Ref. 23

   – – – –  Liquid drop model Ref . 23

**Figure 2.** The deformation energies of $^{234}$U versus the distance between mass centres r.

  The energies of the system using crevice shape parametrization :

   (a) : with finite range and finite surface diffuseness

   (b) : without finite range and finite surface diffuseness

**Figure 3.** Variation of the fission barrier versus the angular momentum for A = 50, 100 and 200 β-stable nuclei.

**Figure 4.** Limit of the nucleus stability due to the centrifugal effects

The maximum angular momentum $L_{BF}$ beyond witch the fission barrier vanish as a function of mass number A for β-stable nuclei

|         |         |
|---------|---------|
| ——      | this Work |
| • • • • • | Ref. 37 |
| – · – · – · | Ref. 35 |
| · · · · · · | Ref. 30 |

**Figure 5.** Region of the hyperdeformation for β–stable nuclei predicted by our calculations

**Figure 6.** The fission path for the $^{48}$Cr calculated with an axial creviced shape ($\alpha = 1$)
For L = 35 $\hbar$.

**Figure 7.** The fission path for the $^{48}$Cr calculated with an axial creviced shape ($\alpha = 1$)
for L = 36 $\hbar$.

**Figure 8.** Energy contours $E_d(s_1, \alpha)$ – the fission path – for the $^{48}$Cr for L = 35 $\hbar$ calculated with a triaxial crevaced shape ($\alpha \neq 1$)

**Figure 9**. Energy contours $E_d(s_1, \alpha)$ – the fission path – for the $^{48}$Cr for L = 36 $\hbar$ calculated with a triaxial crevaced shape ($\alpha \neq 1$)

**Figure 10.** Energy contours $E_d(s_1, \alpha)$ – the fission path – for the $^{48}$Cr for L = 60 $\hbar$ calculated with a triaxial crevaced shape ($\alpha \neq 1$)

**Figure 11.** Fission path for the $^{56}$Ni calculated with for L = 0 $\hbar$

**Figure 12.** Fission path for the $^{56}$Ni calculated with for L = 42 $\hbar$

**Figure 13.** Energy contours $E_d(s_1, \alpha)$ for the $^{56}$Ni calculated with a triaxial crevaced shape ($\alpha \neq 1$) for L = 42 $\hbar$.

**Figure 14.** Predict hyperdeformed states for $20 \hbar < L < 50 \hbar$ for the $^{56}$Ni

**Figure 15.** The fission path for the $^{80}$Zr calculated with an axial creviced shape ($\alpha=1$)
for L=59 $\hbar$

**Figure 16.** Energy contours $E_d(s_1, \alpha)$ for the $^{80}$Zr for L = 59 $\hbar$ calculated with a triaxial crevaced shape ($\alpha \neq 1$).

**Figure 17.** Predicted hyperdeformed states for $20 \hbar < L < 50 \hbar$ the $^{80}$Zr nucleus.

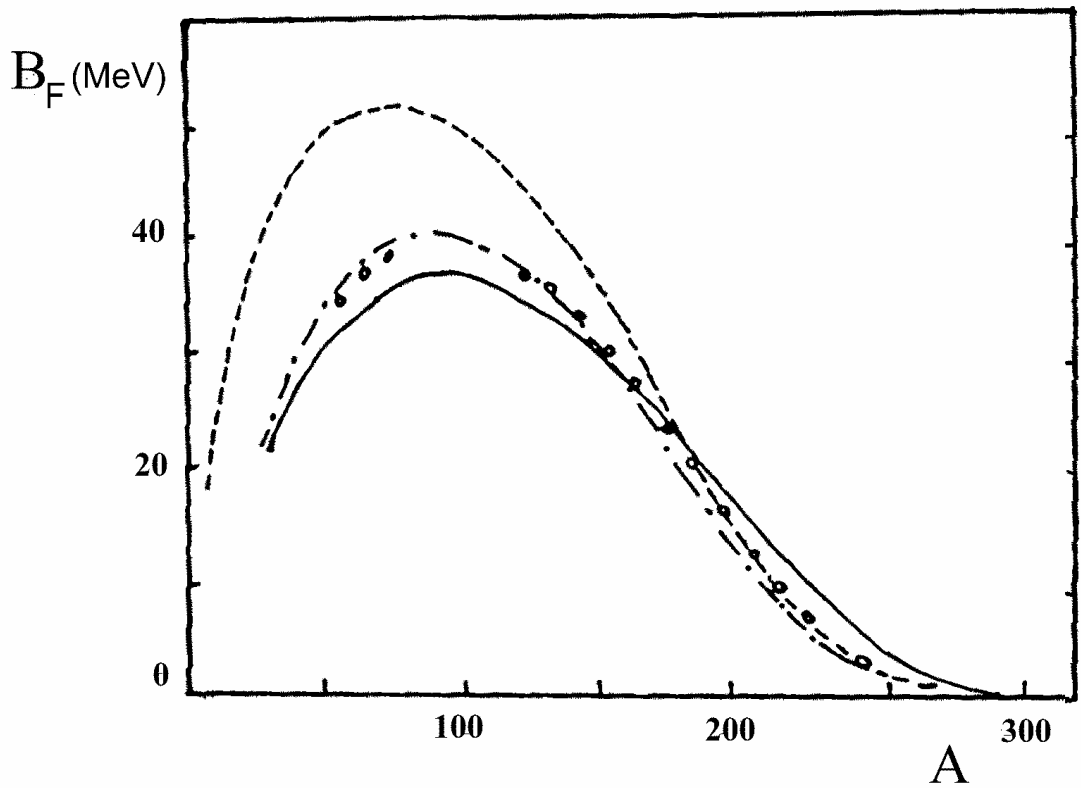

Figure 1

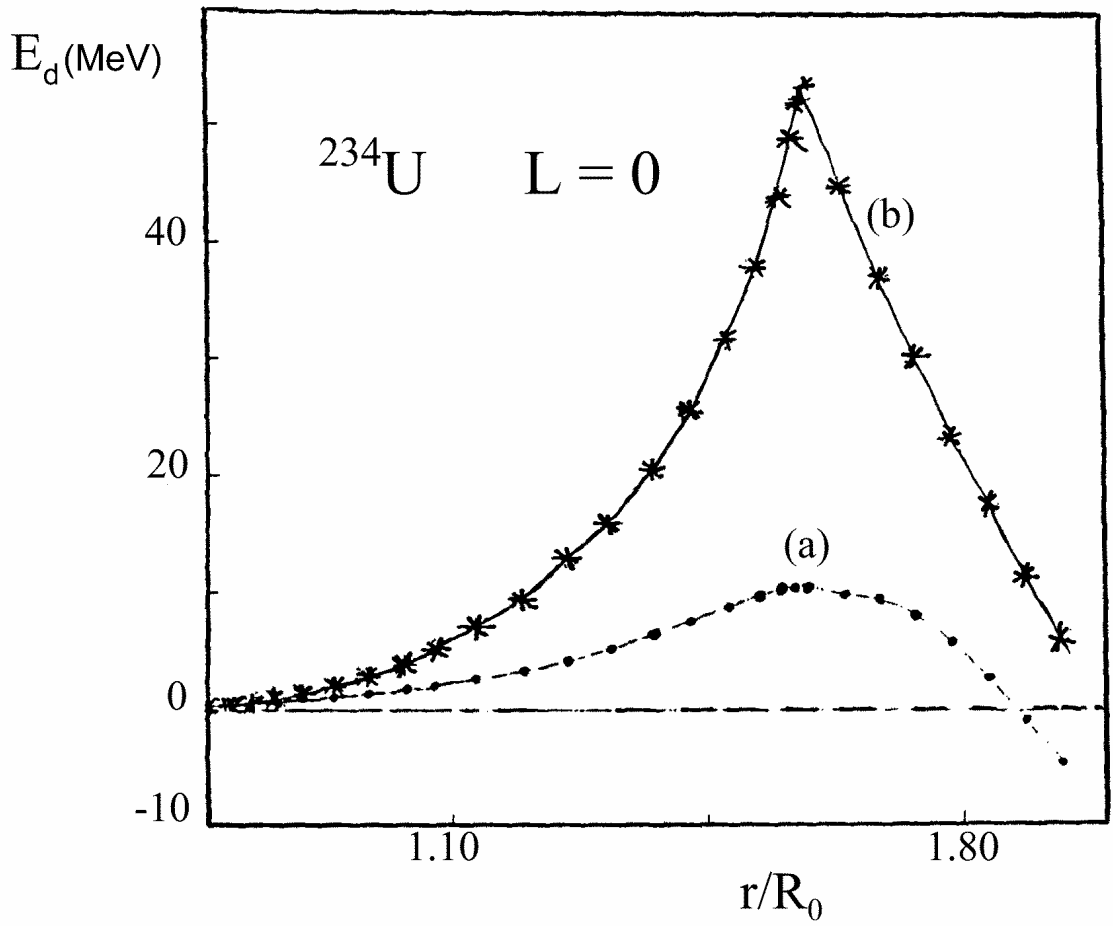

Figure 2

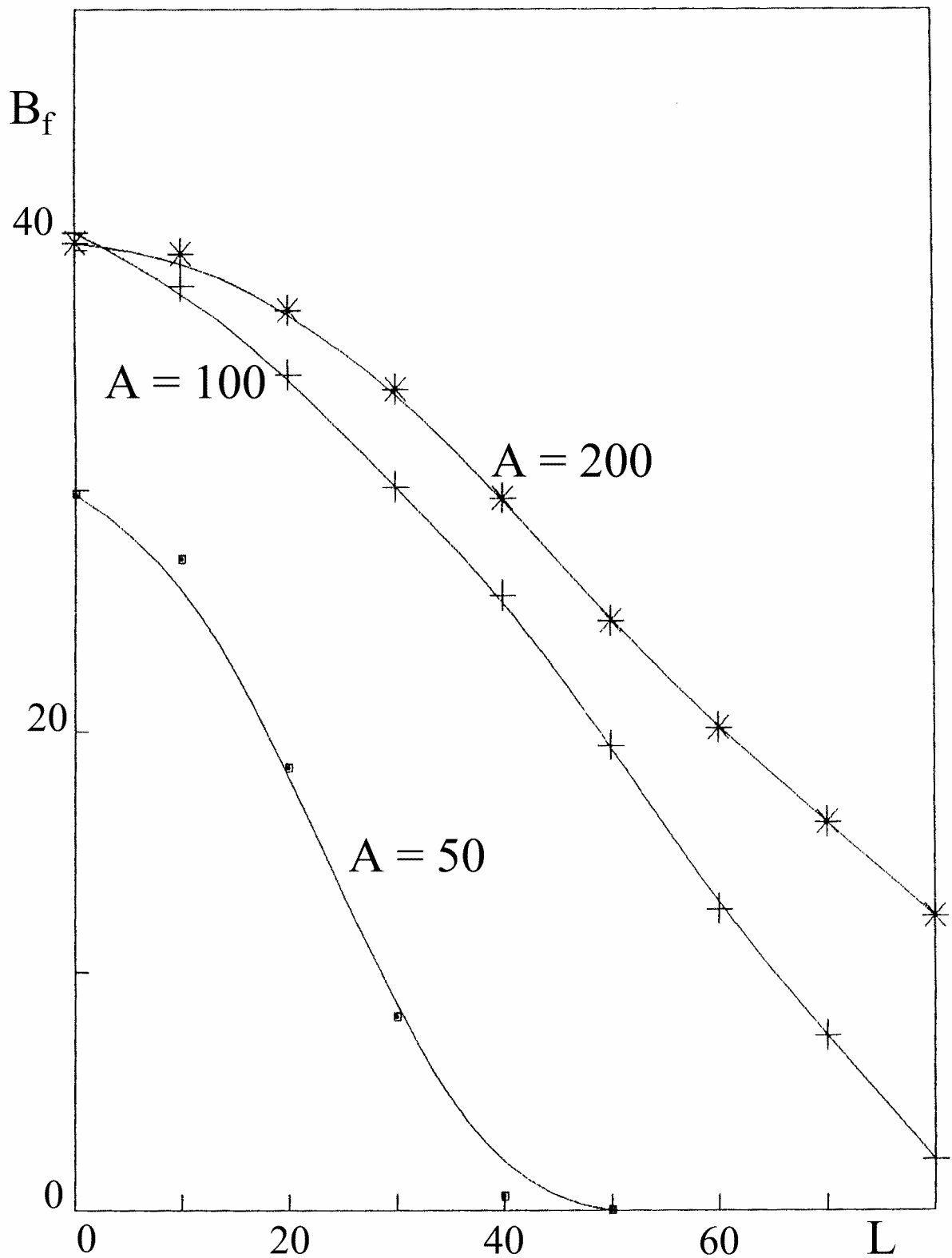

Figure 3

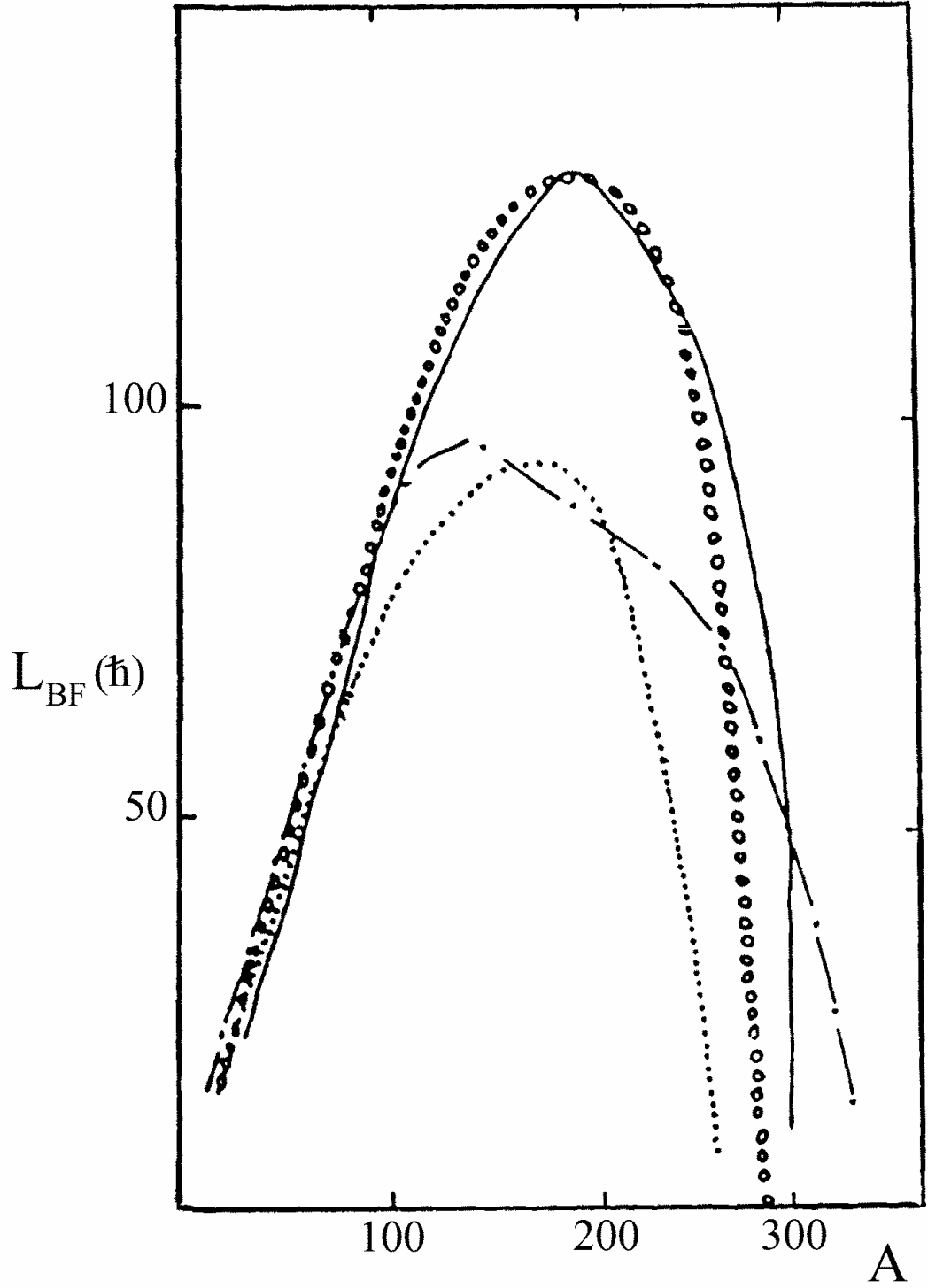

Figure 4

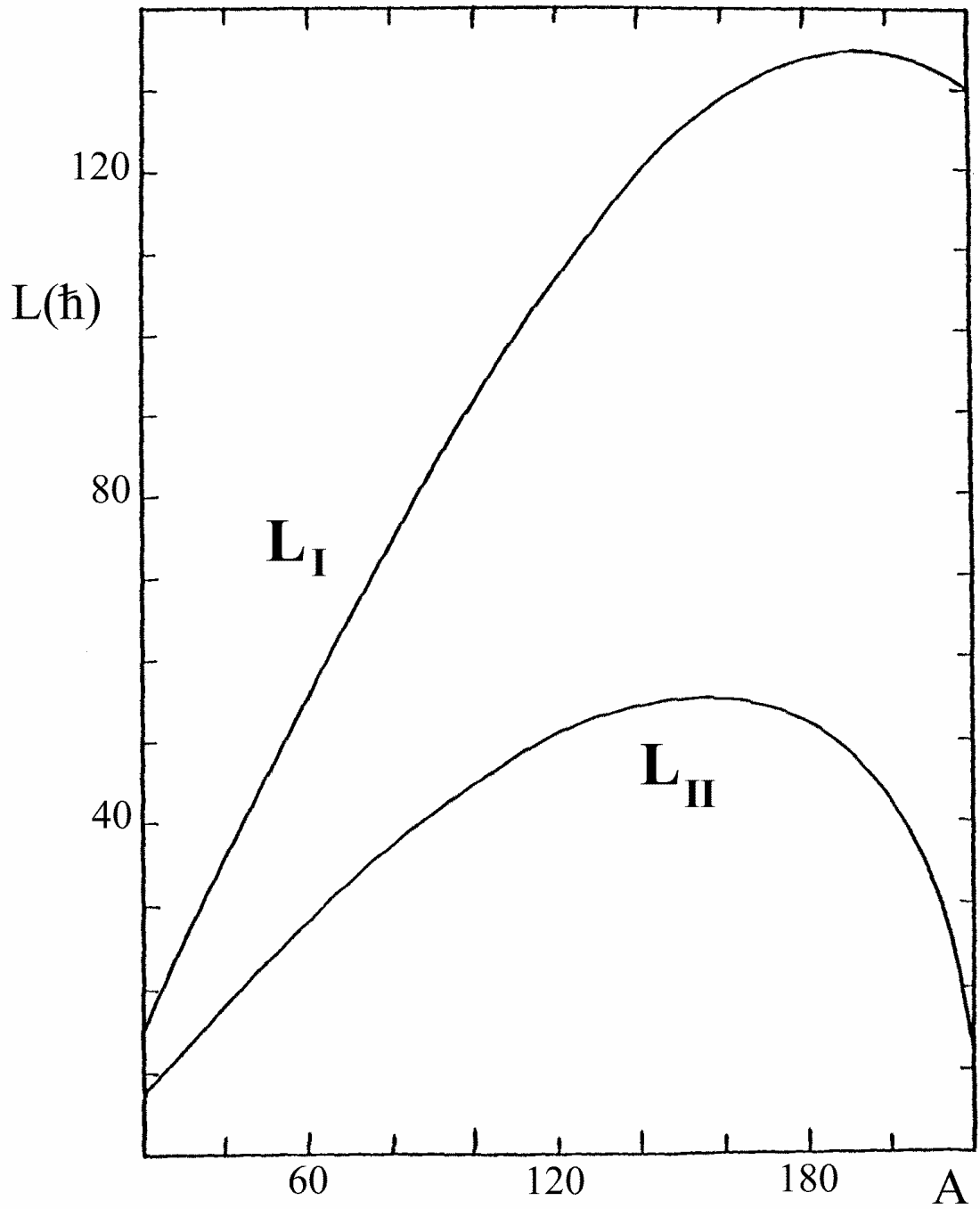

Figure 5

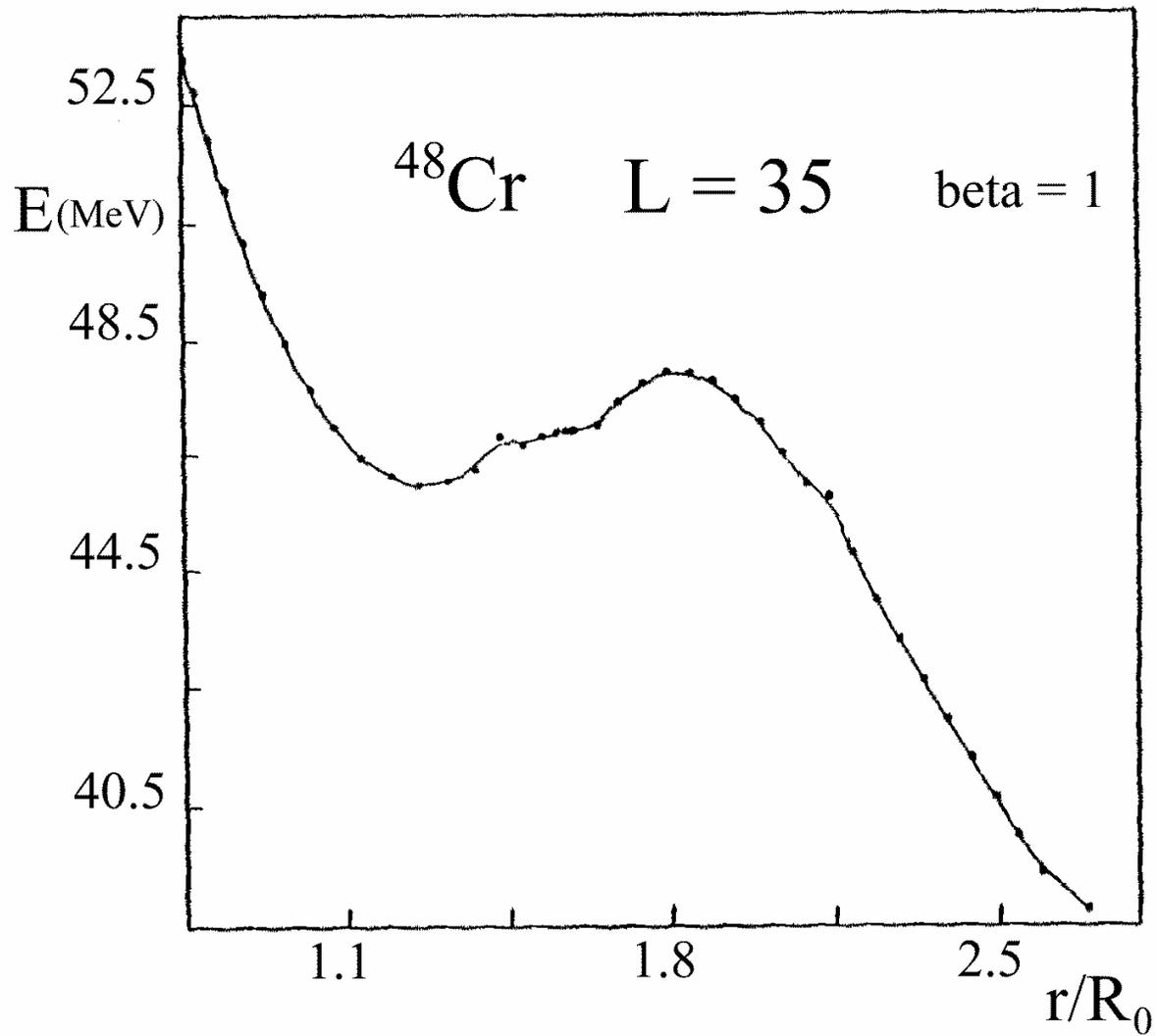

Figure 6

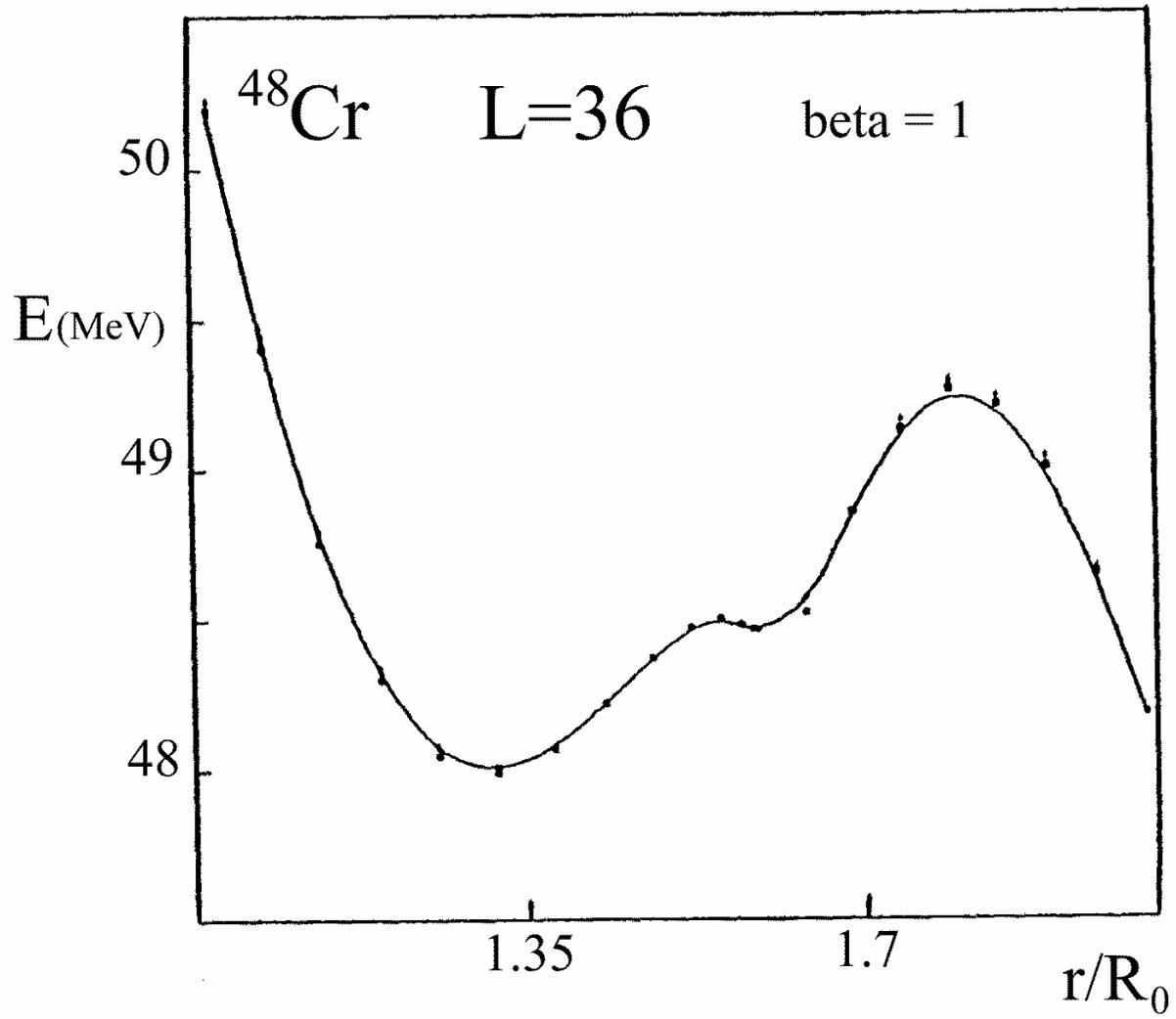

Figure 7

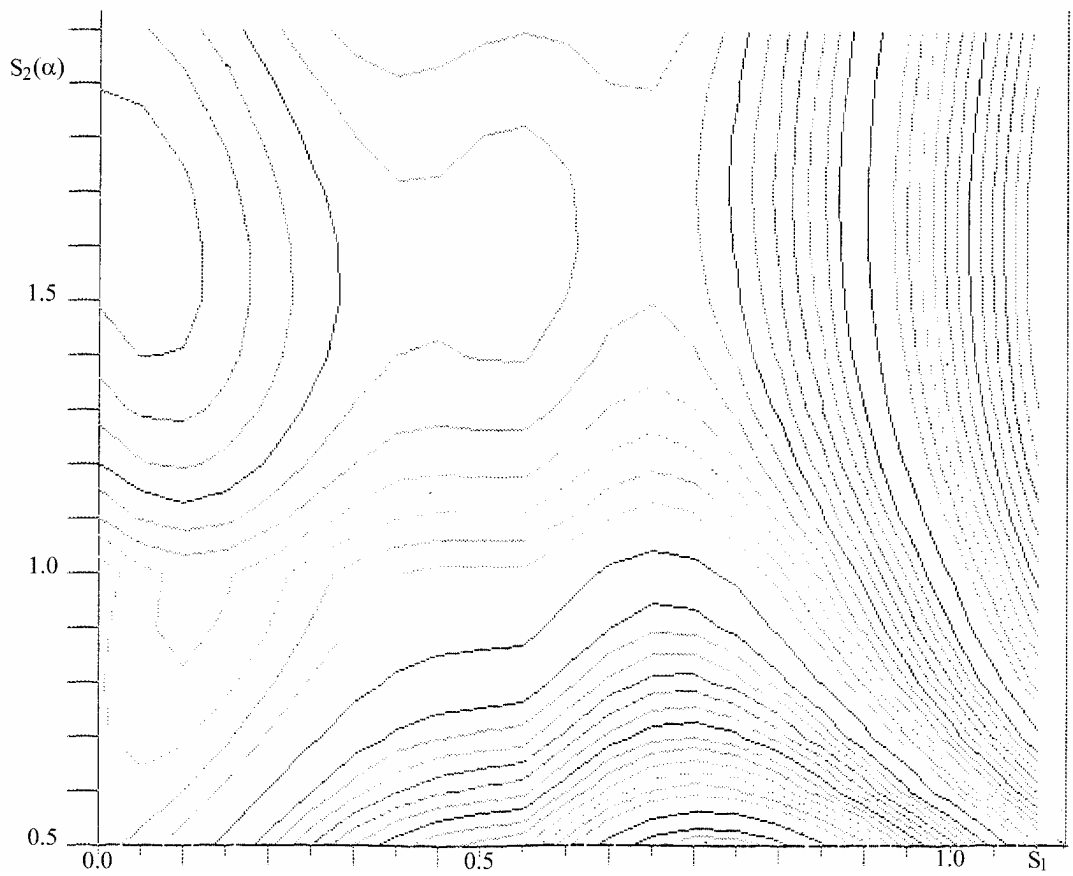

Figure 8

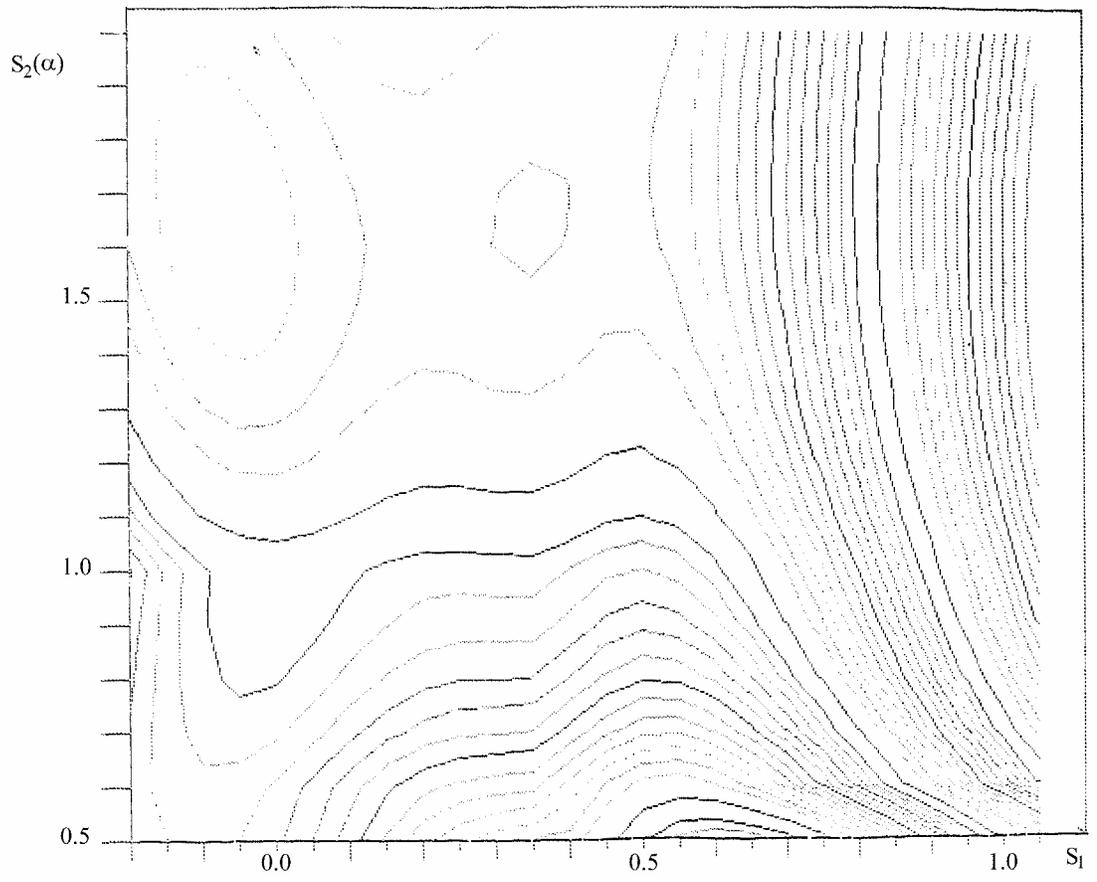

Figure 9

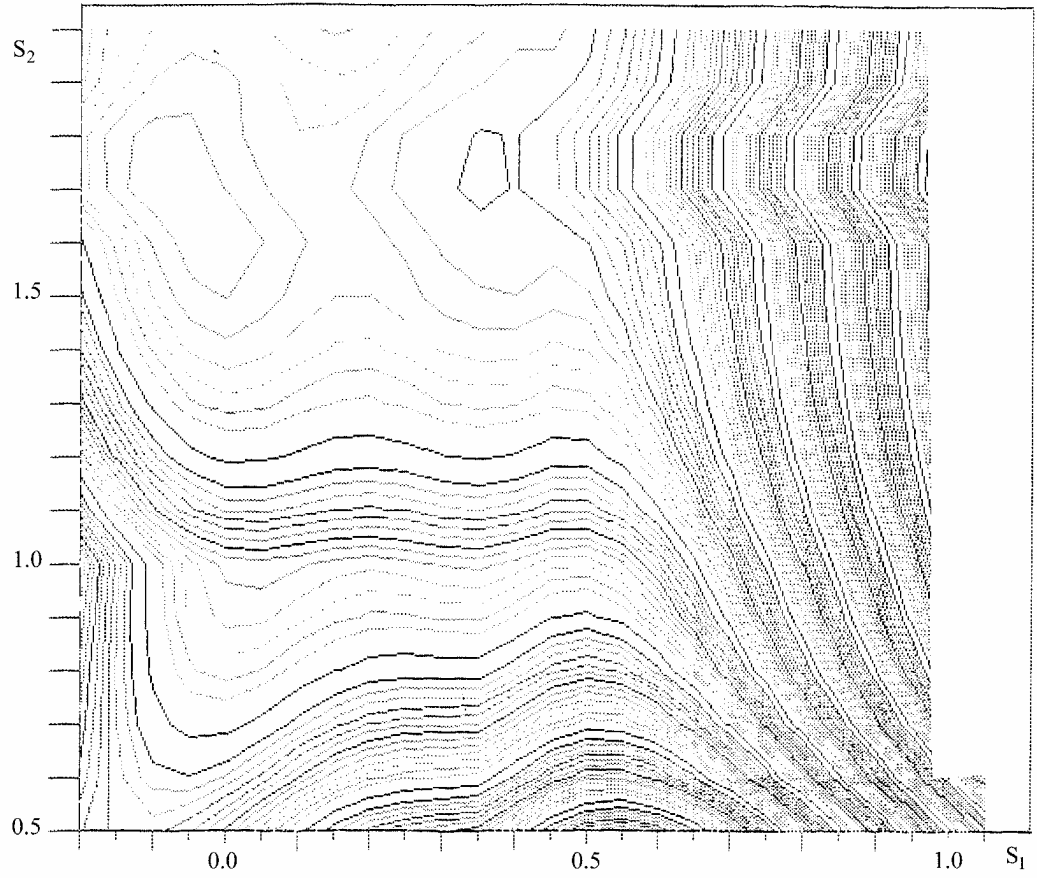

Figure 10

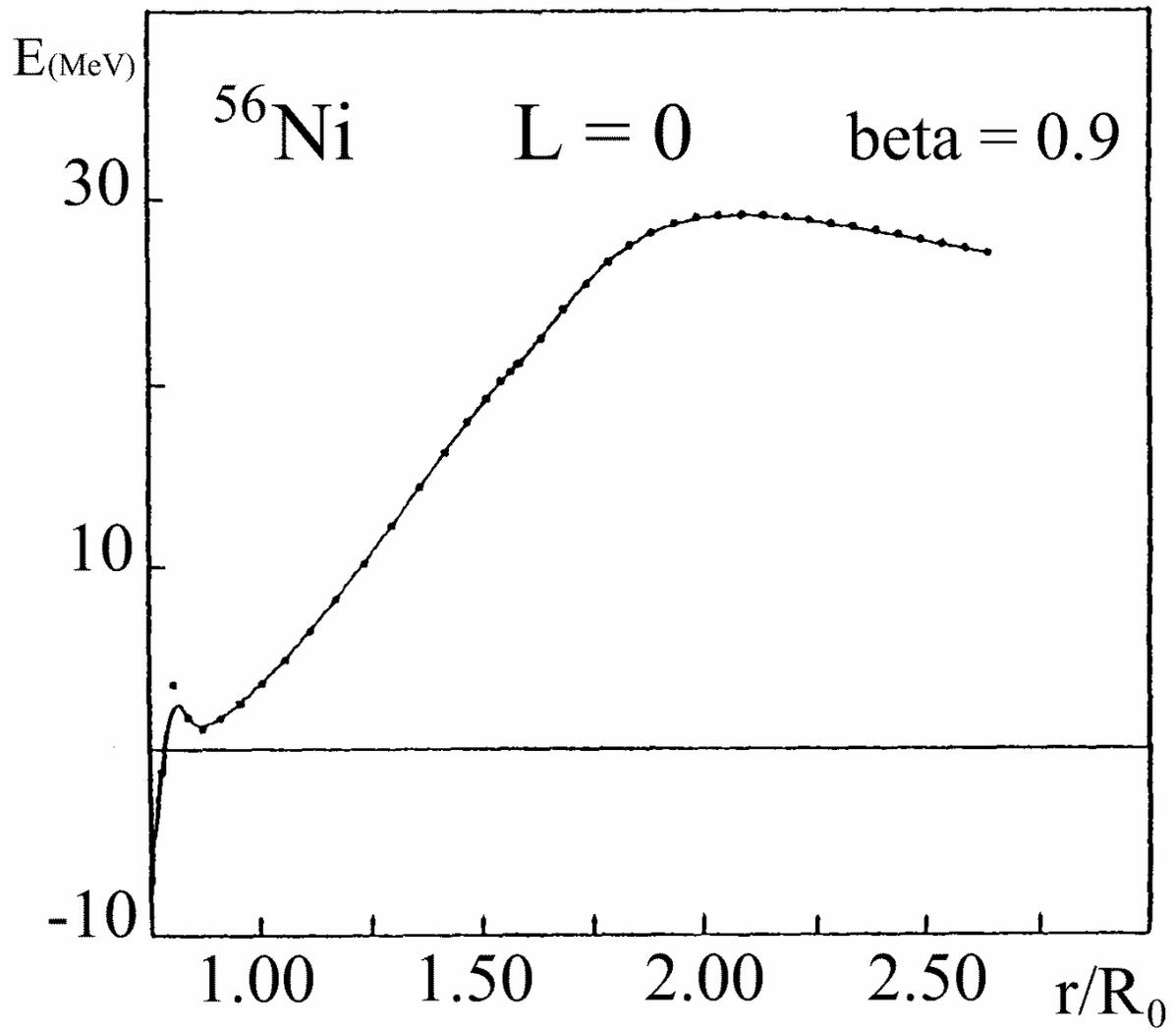

Figure 11

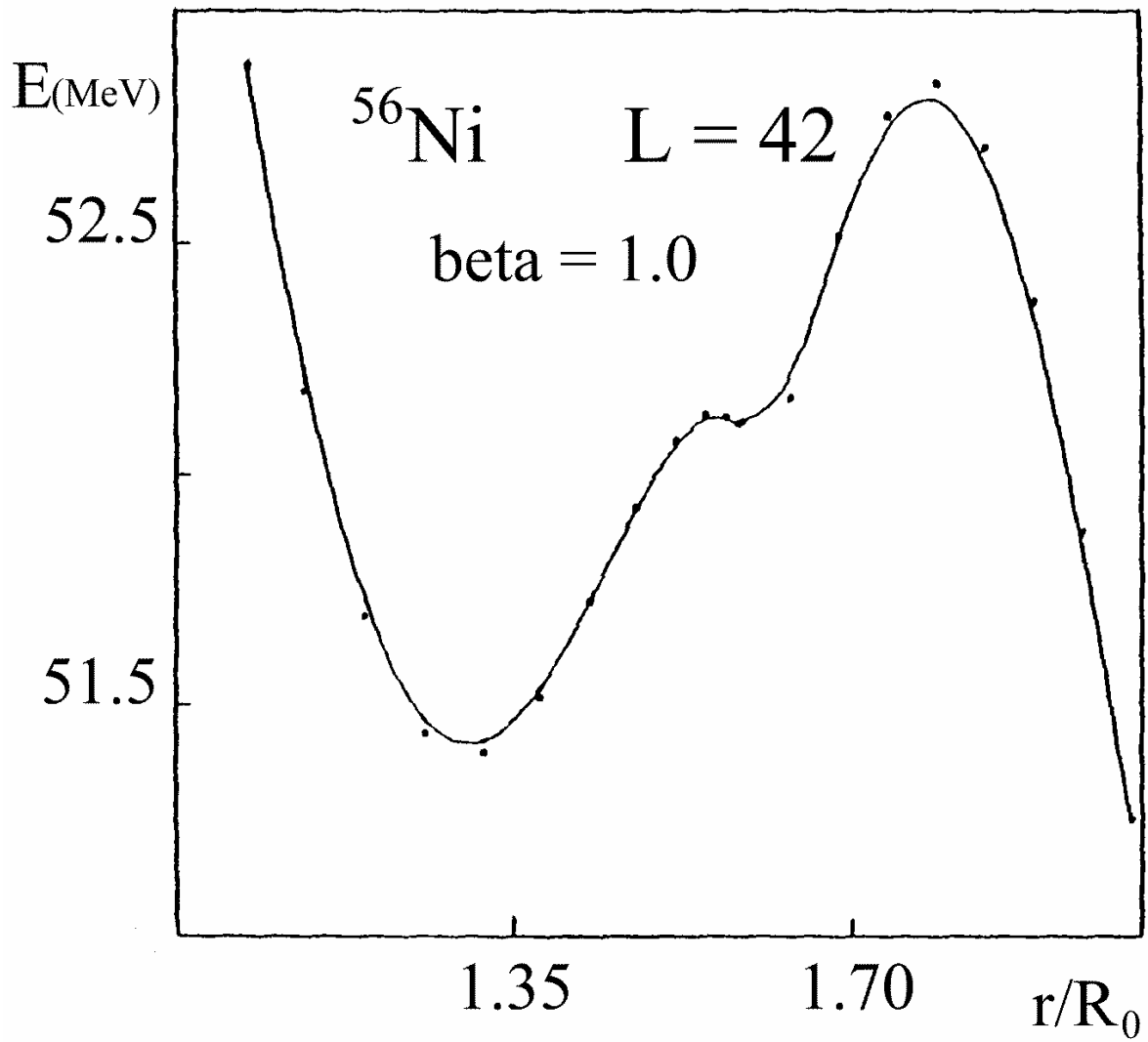

Figure 12

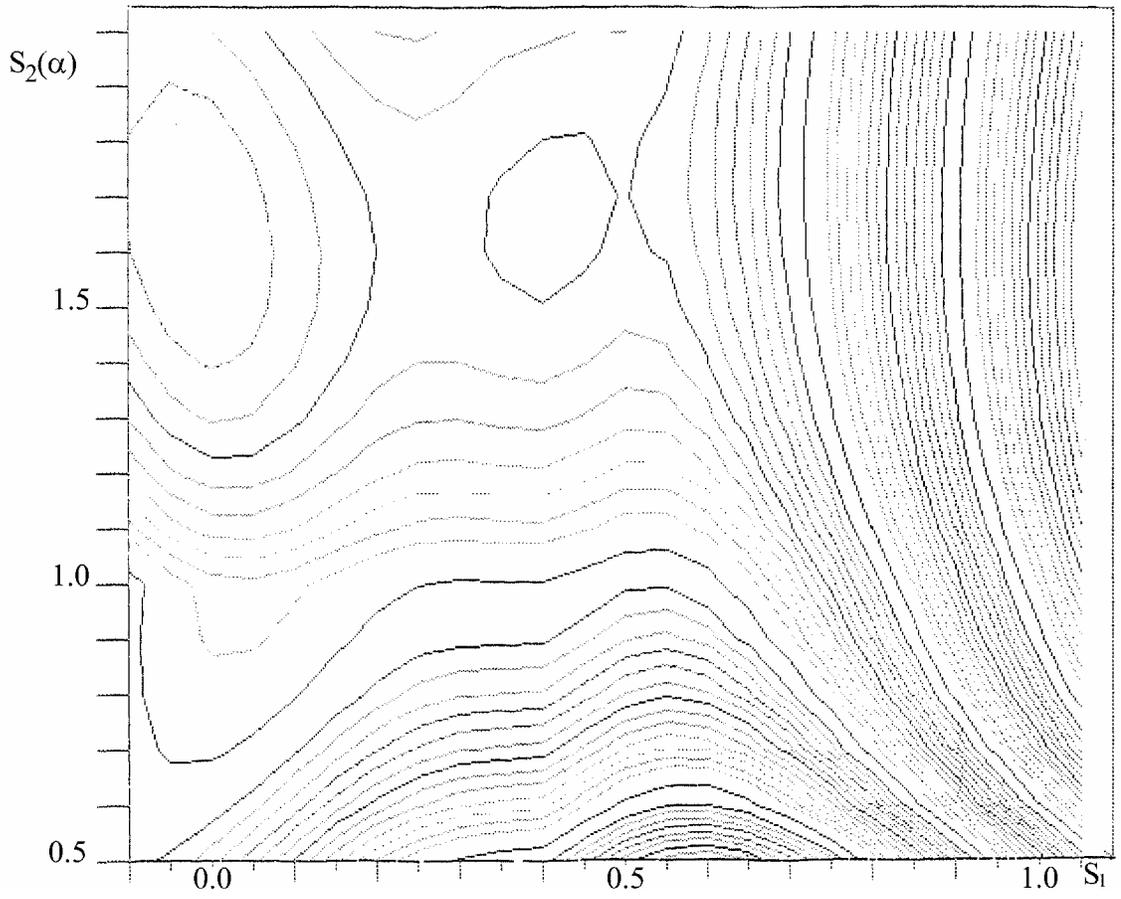

Figure 13

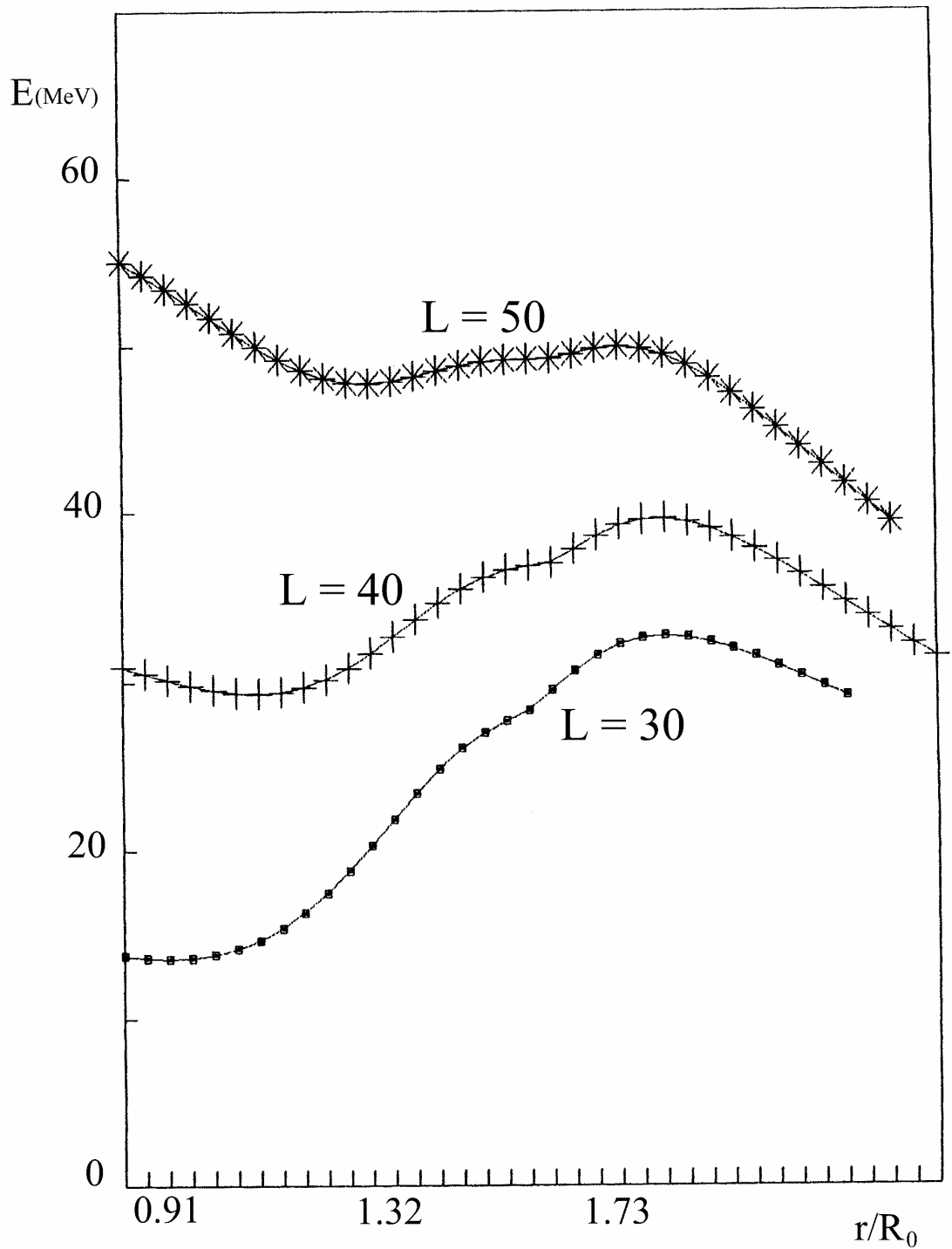

Figure 14

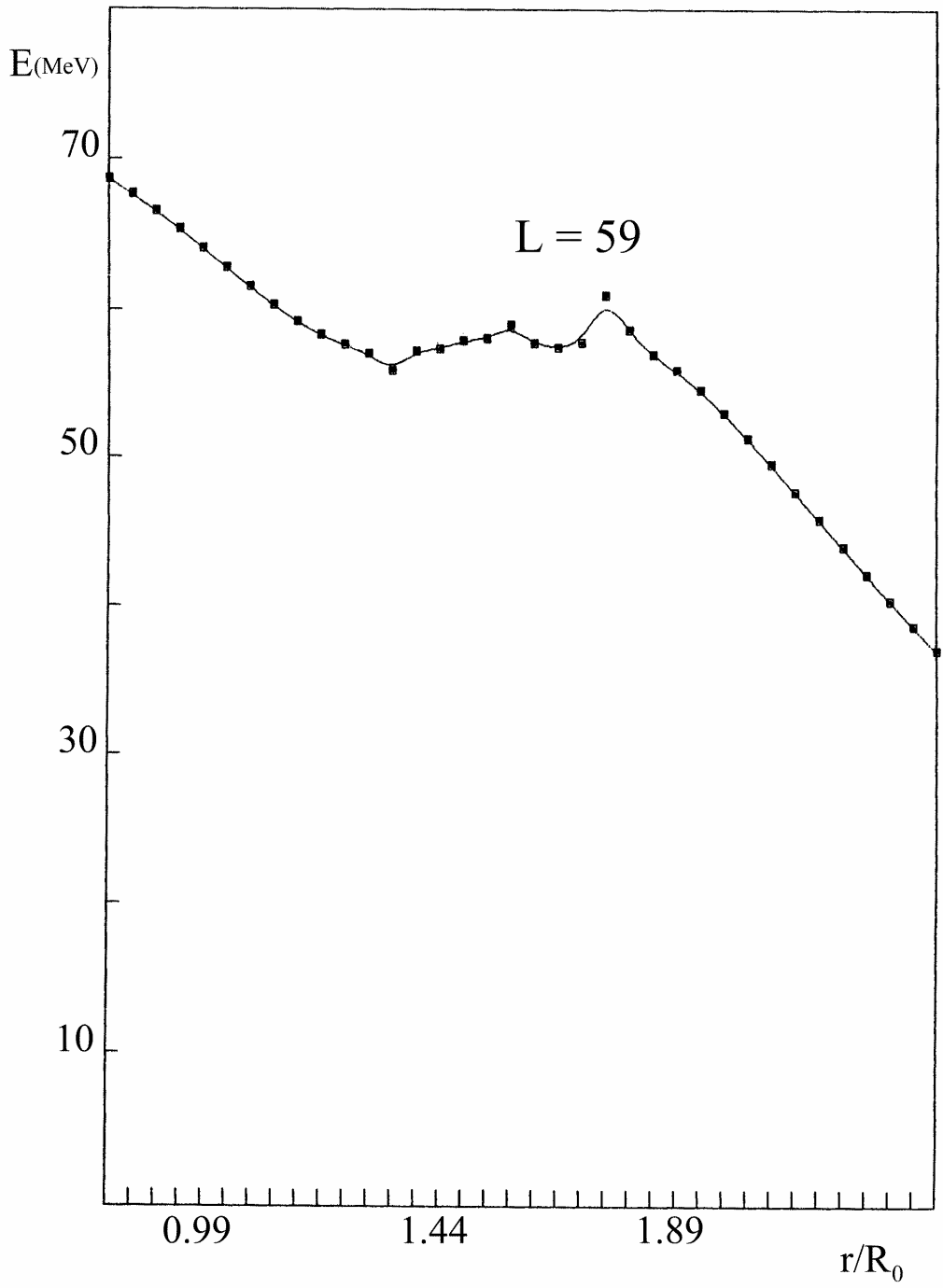

Figure 15

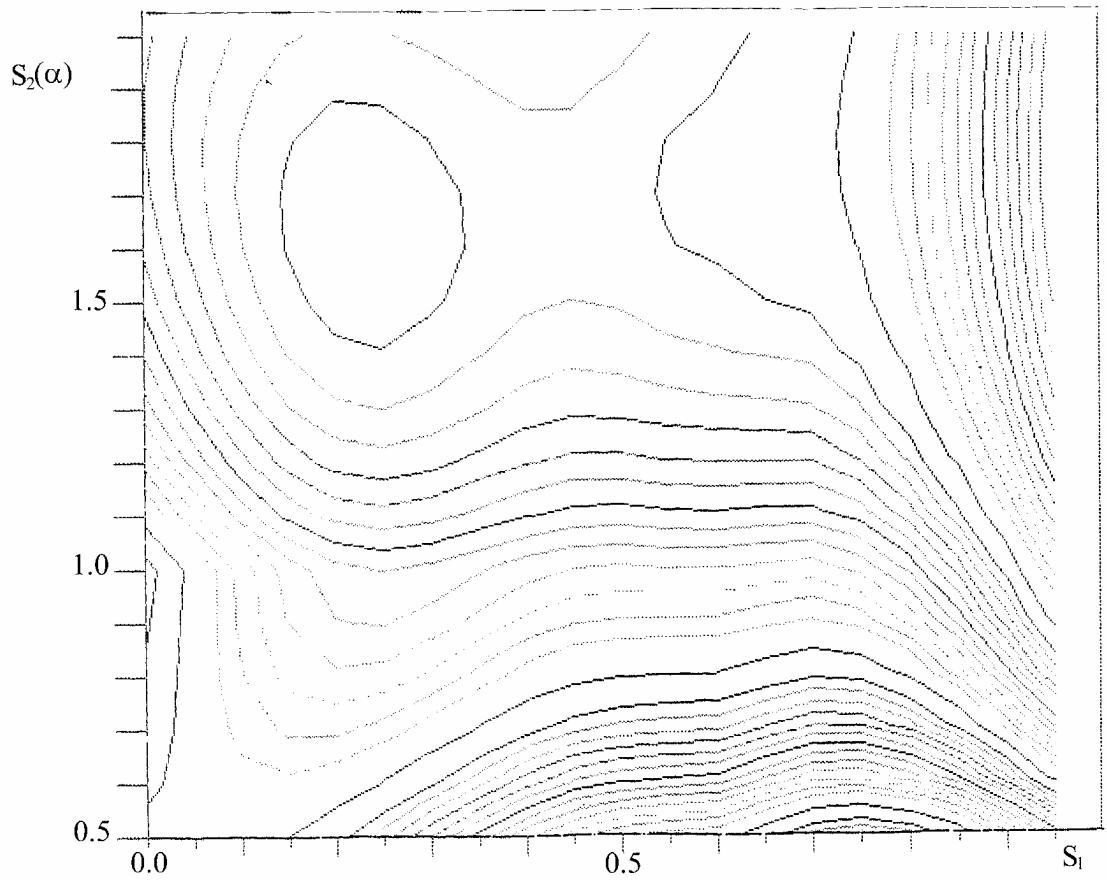

Figure 16

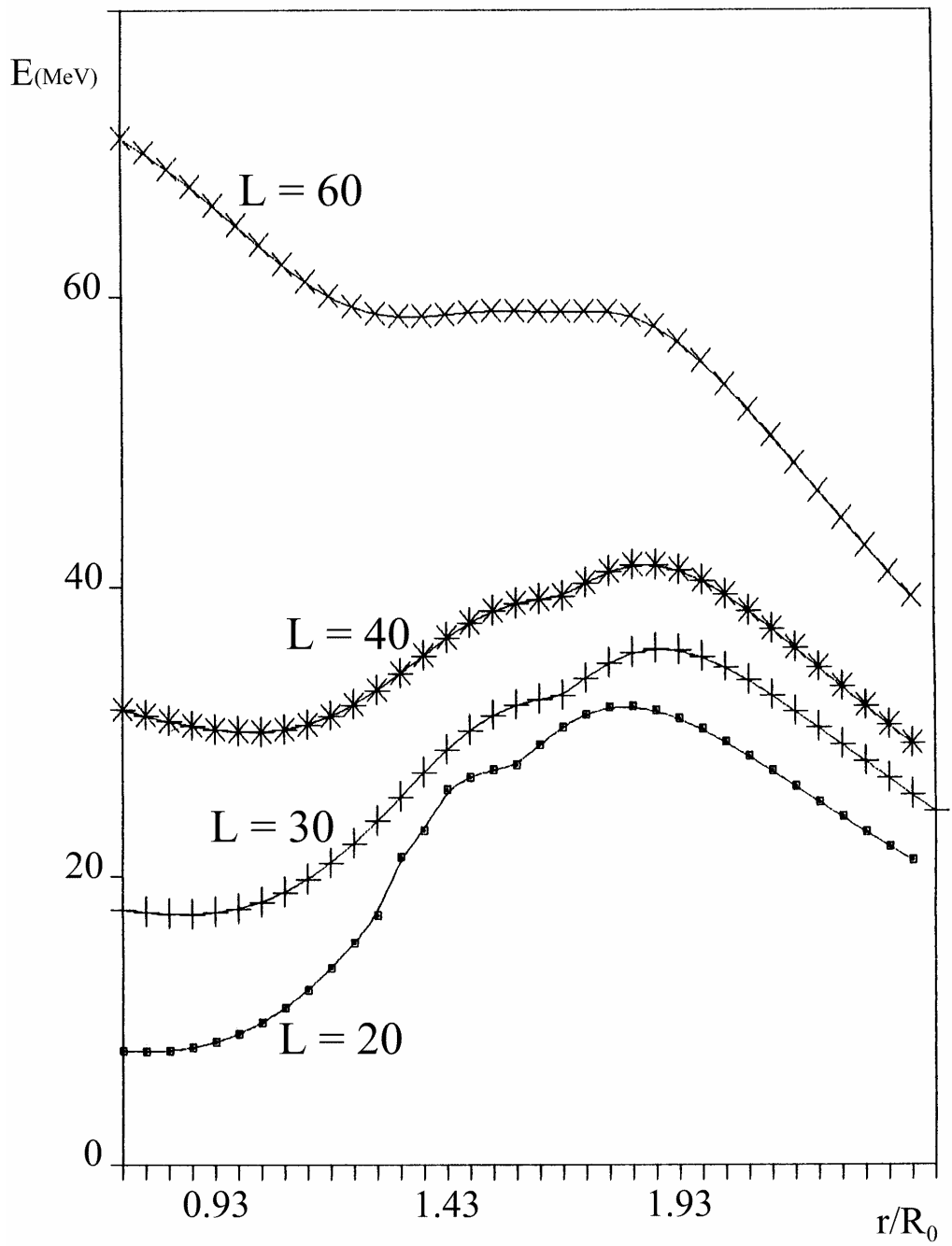

Figure 17